\documentclass{iau} 
\usepackage{graphicx}
\usepackage[export]{adjustbox}
\usepackage{subcaption}

\title[Herschel observations of PNe] %% give here short title %%
{Herschel observations of planetary nebulae 
\thanks{{\it Herschel} is an ESA space observatory with science instruments provided by European-led 
Principal Investigator consortia and with important participation from NASA.}}

\author[G. C. Van de Steene] %% give here short author list %%
{Griet C. Van de Steene$^1$}

\affiliation{$^1$ Royal Observatory of Belgium, Department Astronomy and Astrophysics \\ Ringlaan 3, BE-1180, Brussels, Belgium \\ email: {\tt g.vandesteene@oma.be} \\ 
}

\pubyear{2016}
\volume{323} 
\jname{Planetary nebulae: Multiwavelength probes of stellar and galactic evolution}
\editors{X. Liu, L. Stanghellini, \& A. Karakas, eds.}
\begin{document}

\maketitle

\begin{abstract}
This article presents an overview of the published results for
planetary nebulae based on images and spectroscopy from the PACS,
SPIRE, and HIFI instruments on board the Herschel satellite.
\end{abstract}

\firstsection 
\section{Introduction}

Grains play an important role in many environments, including
planetary nebulae (PNe), because of extinction, photoelectric heating,
their influence on the charge and ionization balance of the gas, as
catalysts for grain-surface chemical reactions and as seeds for
freeze-out of molecules. Previous satellite missions such as IRAS,
ISO, Spitzer, and AKARI have allowed us to study the dust in PNe, but
unfortunately the angular resolution of these instruments was too low
to get detailed information on the spatial distribution of the
dust. This has changed with the Herschel satellite, which has allowed
us to study the spatial structures in unprecedented detail.

\section{Herschel and its instruments}

The Herschel Space Observatory (\cite[Pilbratt et
  al. 2010]{Pilbratt10}) was launched on May 14 2009 and operated for
nearly four years. It carried the largest, most powerful infrared
telescope ever flown in space and three sensitive scientific
instruments. Herschel’s observations finished on April 29 2014 when
the tank of liquid helium used to cool the instruments finally ran
dry.

The three instruments on board were: PACS (Photoconductor Array Camera
and Spectrometer), SPIRE (Spectral and Photometric Imaging REceiver),
and HIFI (Heterodyne Instrument for the Far Infrared), a
high-resolution spectrometer.  These instruments were designed for
deep, wideband photometry with high spatial resolution and full
spectral coverage, making Herschel the first space facility to
completely cover the far infrared and submillimeter range from 55 to
672\,$\mu$m.

PACS (Photodetecting Array Camera and Spectrometer) (\cite[Poglitsch et
al. 2010]{Poglitsch10}) was an imaging camera and low-resolution spectrometer
covering wavelengths from 55 to 210\,$\mu$m. The spectrometer had
a spectral resolution between R\,$=$\,1000 and R\,$=$\,5000. It operated as an
integral field spectrograph, combining spatial and spectral
resolution. The imaging camera was able to image simultaneously in two
bands (either 60 $-$ 85 / 85 $-$ 130\,$\mu$m and 130 $-$ 210\,$\mu$m) with a
detection limit of a few mJy.

SPIRE (Spectral and Photometric Imaging Receiver) (\cite[Griffin et
al. 2010]{Griffin10}) was an imaging camera and low-resolution spectrometer
covering 194 to 672\,$\mu$m wavelength. The spectrometer had a
resolution between R\,=\,40 and R\,=\,1000 at a wavelength of
250\,$\mu$m and was able to image point sources with brightnesses
around 100\,mJy and extended sources with brightnesses of around
500\,mJy. The imaging camera observed simultaneously in three bands,
centred at 250, 350 and 500\,$\mu$m, each with 139, 88 and 43 pixels
respectively. It was able to detect point sources with brightness
above 2\,mJy.

HIFI (Heterodyne Instrument for the Far Infrared) (\cite[de Grauw et
al. 2010]{deGrauw10}) is a heterodyne detector able to electronically separate
radiation of different wavelengths, giving a spectral resolution up to
 R\,=\,10$^7$ . The spectrometer was operated within two wavelength
bands, from 157 to 212\,$\mu$m and from 240 to 625\,$\mu$m.

\section{PACS and SPIRE imaging results}

The extended circumstellar envelopes of evolved low-mass AGB stars
display a large variety of morphologies. Understanding the various
mechanisms that give rise to these extended structures is important to
trace their mass-loss history.  The data presented by \cite[Cox et
  al. (2012)]{Cox12} showed for the first time the variety of
interaction between the circumstellar shell and the interstellar
medium, which can be divided in roughly four categories: ``fermata'',
``eyes'', ``irregular'', and ``rings''.  In particular the star's
peculiar space velocity and the density of the ISM appear decisive in
detecting emission from bow shocks or detached rings. Tentatively, the
"eyes" class objects are associated to (visual) binaries, while the
"rings" generally do not appear to occur for M-type stars, only for C
or S-type objects that have experienced a thermal pulse.  The
occurrence of the observed eye-shape of AGB detached shells is most
strongly influenced by the interstellar magnetic field, the stellar
space motion, and density of the interstellar medium (\cite[van Marle
  et al. 2014]{vanMarle14}).  Observability of this transient phase is
favoured for lines-of-sight perpendicular to the interstellar magnetic
field direction. The simulations of van \cite[Marle et
  al. (2014)]{vanMarle14} indicate that such ``eye'' shapes of such
pre-PN circumstellar shell can strongly affect the shape and size of
PNe.

A total of 18 well known PNe have been imaged with PACS and SPIRE
instruments in the framework of MESS and HerPlans programs.  Seven PNe
(NGC~6720, NGC~650, NGC~7293, NGC~6853, NGC~3587, NGC~7027) were
imaged by the MESS team. Mass loss of Evolved StarS (MESS) was a
Guarenteed Time Key Programmes to study the circumstellar environment
of evolved post main sequence stars. A detailed description of the
program can be found in \cite[Groenewegen et
  al. (2011)]{Groenewegen11}. An overview of the Herschel observations
of PNe in the MESS program was presented in \cite[van Hoof et
  al. (2012)]{vanHoof12}. Eleven PNe (NGC~40, NGC~2392, NGC~3242,
NGC~6445, NGC~6543, NGC~6720, NGC~6781, NGC~6826, NGC~7009, NGC~7026,
Mz~3) have been observed in the Herschel Planetary Nebulae Survey
(HerPlanS). A data overview and first analysis was presented in
\cite[Ueta et al. (2014)]{Ueta14}. HerPlaNS obtained far-infrared
broadband images and spectra of eleven well-known PNe with the PACS
and SPIRE instruments. The target PNe all have distances less than
$\sim$1.5 kpc and are dominated by relatively high-excitation nebulae
as they were selected from the Chandra Planetary Nebula Survey
(ChanPlaNS; \cite[Kastner et al. 2012]{Kastner12})

Herschel PACS and SPIRE imaging showed that the dust emission in PNe
has a very clumpy structure for all nebula. There is excellent
agreement between the H$_2$ images of and the PACS 70\,$\mu$m
maps. For the Ring nebula (NGC\,6720) it appears to be the first
observational evidence that H$_2$ forms on oxygen rich dust grains.
\cite[van Hoof et al. (2010)]{vanHoof10} developed a photoionisation
model of the Ring nebula with {\tt Cloudy} to investigate possible formation
scenarios for H$_2$. They concluded that the most plausible scenario
is that the H$_2$ resides in high density knots which were formed
after the recombination of the gas started, when the central star
luminosity dropped steeply after the central star entered the cooling
track. H$_2$ formation may still be ongoing at this moment, depending
on the density of the knots and the properties of the grains in the
knots (\cite[van Hoof et al. 2010]{vanHoof10}). This is also a
possible scenario for the formation of high density clumps in other
evolved nebula with a central star on the cooling track such as the
Helix (NGC~7293) and the Dumbell (NGC~6853).

Comparison between the 70\,$\mu$m Herschel and corresponding optical
maps showed that they are very similar indicating that there is a very
steep temperature gradient from the ionized region to the dusty
photodissociation region.  For NGC\,6781 the PACS 70\,$\mu$m map,
showing the distribution of thermal dust continuum is very similar to
what is seen in the [N{\sc ii}]$\lambda$658.4 nm image (\cite[Ueta et
  al. 2014]{Ueta14}). For the Helix it was also observed and shown
that the radiation field decreases rapidly outwards through the barrel
wall (Fig. 9, \cite[Van de Steene et al. 2015]{VandeSteene15}).

Previous knowledge of the 3D structure of the nebula is extremely
useful to correctly interpret the far infrared images. For instance,
the PACS and SPIRE images of the Helix nebula could be understood
based on the kinematic model of \cite[Zeigler et
  al. (2013)]{Zeigler13} and the Herschel images of NGC~6781 with the
3D model of \cite[Schwarz \& Monteiro (2006)]{Schwarz06}.  Both show
bipolar, barrel-like structures inclined to the line of sight, a
frequent morphology in PNe.

Herschel PACS imaging photometry was obtained for these 17 different
PNe in the MESS and HerPlanS projects (\cite[van Hoof et
  al. 2010]{vanHoof10}, \cite[van Hoof et al. 2013]{vanHoof13},
\cite[Van de Steene et al. 2015]{VandeSteene15},Van de Steene et
al. in preparation, \cite[Ueta et al. 2014]{Ueta14}, Ueta et al. in
preparation) with the PACS and SPIRE instruments at 70, 160, 250, 350,
and 500\,$\mu$m.  This photometry was complemented with photometry
obtained from the literature at many other wavelengths from the UV to
radio wavelengths to construct full SEDs. The modified black body fit
to these SEDs revealed that the emission factor $\beta$ is always
close to 1.0, indicating that the dust grains are mainly amorphous
carbon (\cite[Menella et al. 1995]{Menella95}, \cite[Boudet et
  al. 2005]{Boudet05}).  The fit to the SED also showed that the flux
emitted in the far infrared is significant: without far-IR data
fitting constraints the dust mass gets underestimated by 40\%.

The dust temperature obtained from the SED fits and the temperature
maps made, showed that the cool dust temperature of the PNe is around
30 to 100\,K.  For the Helix nebula the gas kinetic temperature T$_k$
was determined to be about 20 to 40~K (\cite[Zack \& Ziurys
  2013]{Zack13}, \cite[Etxaluze et al. 2014]{Etxaluze14}), which is
similar the Helix' dust temperature (\cite[Van de Steene et
  al. 2015]{VandeSteene15}). The gas density of the H$_2$ cometary
globules is on the order of n(H$_2$)\,$\sim$\,(1–5)\,10$^5$\,cm$^{-3}$. 
\cite[Goldsmith (2001)]{Goldsmith01} found
that for gas densities higher than 10$^{4.5}$\,cm$^{-3}$ the dust and
gas temperatures will be closely coupled, also for the dust
temperatures determined for the Helix nebula.

The dust masses found so far for NGC\,6781, NGC\,7293, and NGC\,650
are all a few thousandths of solar masses.  By integrating over the
entire nebula, the dust column mass density map the total mass of far
emitting dust mass was determined to be 4\,x\,10$^{-3}$\,M$_\odot$ for
NGC\,6781 at a distance of 950\,pc and 3.5\,x\,10$^{-3}$\,M$_\odot$
for the Helix nebula at distance of 216\,pc, while for NGC~650 the dust mass
is about 1.4\,x\,10$^{-3}$\,M$_\odot$ at a distance of 1200\,pc based on
Cloudy modeling (\cite[Ueta et al. 2014]{Ueta14}, \cite[Van de Steene et al. 2015]{VandeSteene15}, \cite[van Hoof et al. 2013]{vanHoof13})

One of the goals of HerPlaNS is to empirically obtain spatially
resolved gas-to-dust mass ratio distribution maps by deriving both the
dust and gas column mass distribution maps directly from observational
data.  For NGC\,6781 direct comparison of the dust and gas column mass
maps constrained data allowed to construct an empirical gas-to-dust
mass ratio map, which showed a large range of ratios with the median
of 195$\pm$110 and hence, is generally consistent with the typical
spatially-unresolved ratio between 100 and 400 widely used in the
literature for the case of PNe and AGB stars (\cite[Ueta et al. 2014]{Ueta14}).

The MESS and HerPlaNS teams have collected not only photometry, but
also other spectroscopic data from the literature over the whole
spectral range from X-rays to radio to make the most comprehensive
Cloudy models ever made of NGC\,650 (\cite[van Hoof et al. 2013]{vanHoof13}) and
NGC\,6781 (M. Otsuki, this volume).  For NGC~650 the {\tt Cloudy} model
showed that the grains in the ionized nebula are large (assuming
single-sized grains, they would have a radius of 0.15~$\mu$m). Most
likely these large grains were inherited from the asymptotic giant
branch phase. However the PACS 70$/$160\,$\mu$m temperature map showed
evidence of two radiation components heating the grains. The first
component is direct emission from the central star, while the second
component is diffuse emission from the ionized gas (mainly
Ly$\alpha$). Unlike what was thought before, the neutral material
resides in dense clumps inside the ionized region. These may also
harbour stochastically heated very small grains in addition to the
large grains.  This is unusual for such a highly evolved PN.

In the past, far-IR SED fitting with broadband fluxes were performed
under the assumption of negligible line contamination. With the
Herschel data and {\tt Cloudy} modeling we verified that the degree of line
contamination is approximately 8−20\% (\cite[Ueta et
  al. 2014]{Ueta14}, \cite[van Hoof et al. 2013]{vanHoof13}) and does
not significantly affect the fitting results.

\section{PACS and SPIRE spectroscopic results}

\subsection{NGC 6781}

The Herschel spectra obtained at various locations within NGC\,6781
revealed both the physical and chemical nature of the nebula. The
spectra showed a number of ionic and atomic lines such as [O\,{\sc iii}]\,52, 88\,$\mu$μm, [N\,{\sc iii}]\,57\,$\mu$m,
[N\,{\sc ii}]\,122, 205\,$\mu$m, [C\,{\sc ii}]\,158\,$\mu$m, and
[O\,{\sc i}]\,63, 146\,$\mu$m, as well as various molecular lines, in
particular, high-J CO rotational transitions, OH, and OH$^+$ emission
lines. Thermal dust continuum emission was also detected in most bands
in these deep exposure spectra. On average, the relative distributions
of emission lines of various nature suggested that the barrel cavity
in NGC\,6781 is uniformly highly ionized, with a region of lower
ionization delineating the inner surface of the barrel wall.
The least ionic and atomic gas, molecular, and dust species are
concentrated in the cylindrical barrel structure.

Based on the PACS IFU spectral cube data, \cite[Ueta et
  al. (2014)]{Ueta14} derived line maps in the detected ionic and
atomic fine-structure lines. Next diagnostics of the electron
temperature and density using line ratios such as [O\,{\sc iii}]
52/88\,$\mu$m and [N\,{\sc ii}] 122/205\,$\mu$m, resulted in (T$_e$,
n$_e$) and ionic/elemental/relative abundance profiles for the first
time in the far-IR for any PN.  The derived T$_e$ profile
substantiated the typical assumption of uniform T$_e$ = 10$^4$ K in
the main ionized region, while showing an interesting increase in the
barrel (up to 10\% higher), followed by a sudden tapering off toward
the halo region.  The n$_e$ profile of high-excitation species is
nearly flat across the inner cavity of the nebula, whereas the n$_e$
profile of low excitation species exhibits a radially increasing
tendency with a somewhat complex variation around the barrel wall.  In
fact, this n$_e$[N\,{\sc ii}] profile is reflected in the physical
stratification of the nebula revealed by the ionic/elemental abundance
analysis. The detected stratification is consistent with the previous
inferences made from the past optical imaging observations in various
emission lines of varying levels of excitation.  The derived relative
elemental abundance profiles showed uniformly low N and C abundances,
confirming the low initial mass ($<$ 2 M$_\odot$) and marginally
carbon-rich nature of the central star. However, the profiles did not
appear to reveal variations reflecting the evolutionary change of the
central star, such as a radially increasing carbon abundance.

\subsection{SPIRE spectroscopy: OH$^+$ and CO}

\begin{figure}[ht]
\begin{subfigure}{0.5\textwidth}
\begin{center}
\includegraphics[width=0.9\linewidth, left]{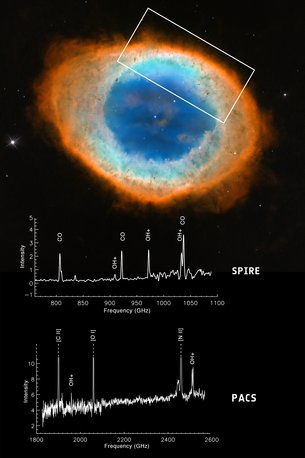}
\caption{SPIRE and PACS spectra of molecules in the Ring nebula NGC\,6720.}
\label{fig:subim1}
\end{center}
\end{subfigure}
\begin{subfigure}{0.5\textwidth}
\begin{center}
\includegraphics[width=0.9\linewidth, right]{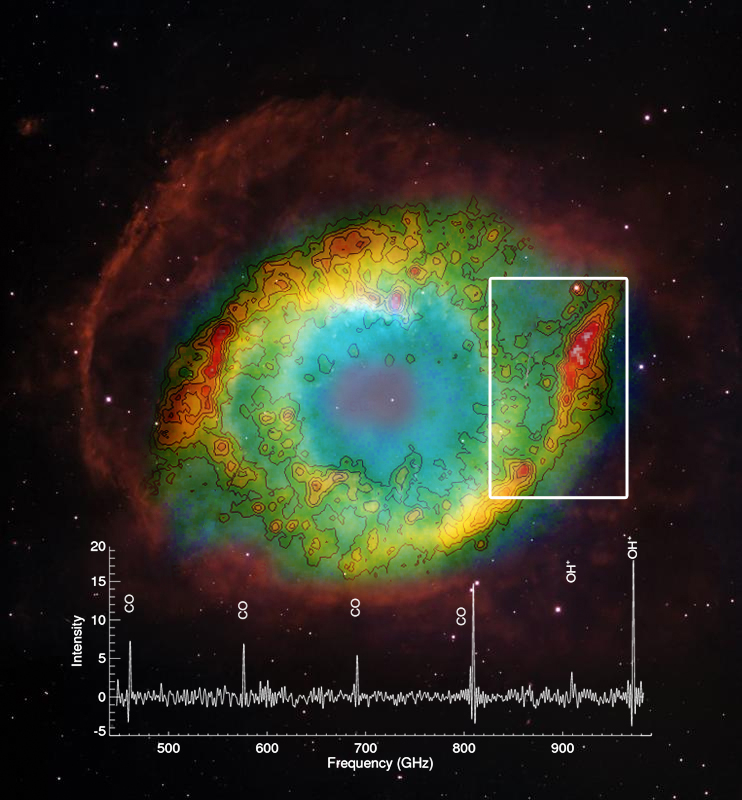}
\caption{SPIRE 250\,$\mu$m image and spectrum of molecules in the Helix nebula NGC\,7293.}
\label{fig:subim2}
\end{center}
\end{subfigure}
\begin{center}
\caption{Herschel spectra of molecules in the Ring and Helix nebula}
\label{fig1}
\end{center}
\end{figure}

\cite[Etxaluze et al. (2014)]{Etxaluse14} and \cite[Aleman et
  al. (2014)]{Aleman14} reported the first detection of extended
OH$^+$ lines in emission in 5 PNe observed as part of the HerPlans in
NGC~6445, NGC~6720 (Fig. \ref{fig:subim1}), and NGC~6781 and MESS in
NGC~7293 (Fig. \ref{fig:subim2}).  Also NGC~6853 shows OH$^+$ in
emission (Van de Steene et al., in preparation).  All five PNe are
molecule rich, with dense clumpy structures and hot central stars
(Teff $>$ 100000\,K).  The OH$^+$ emission is most likely due to
excitation in a photodissociation region. 
Although other factors such as high density and
low C/O ratio may also play a role in the enhancement of the OH$^+$
emission. The fact that OH$^+$ is not detected in objects with
T$_{eff}$ $<$ 100000~K suggests that the hardness of the ionising
central star spectra could be an important factor in the production of
OH$^+$ emission in PNe.

The Herschel spectra towards the Helix nebula also show, besides OH$^+$, CO
emission lines (from J = 4 to 8), [N\,{\sc ii}] at 1461~GHz from
ionized gas, and [C\,{\sc i}] ($^3$P$_2$$-$$^3$P$_1$). The SPIRE spectral maps
suggest that CO arises from dense and shielded clumps in the western
rims of the Helix nebula, whereas OH$^+$ and [C\,{\sc i}] lines trace
the diffuse gas and the UV and X-ray illuminated clump surfaces where
molecules reform after CO photodissociation. The [N\,{\sc ii}] line
traces a more diffuse ionized gas component in the interclump medium
(\cite[Etxaluze et al. 2014]{Etxaluze14}).

For NGC~6781 the CO observations and analysis with higher-J
transitions sampled much warmer CO gas component in the cylindrical
barrel structure, probably located closer to the equatorial region
along the line of sight, compared with the previous CO measurements
and diagnostics by \cite[Bachiller et
  al. (1993)]{Bachiller93}. However, the amount of this warm component
was determined to be an order of magnitude smaller than the cold
component (\cite[Ueta et al. 2014]{Ueta14}).

\subsection{Crystalline olivine}

\cite[Blommaert at al. (2014)]{Blommaert14} (GT1 "Forsterite dust in
the circumstellar environment of evolved stars") presented 48
PACS spectra of evolved stars in the wavelength range of
67\,$-$\,72 $\mu$m, covering the 69\,$\mu$m band of crystalline olivine
(Mg$_{2−2x}$Fe($_{2x)}$SiO$_4$).  For 27 objects in the sample, they
detected the 69\,$\mu$m band of crystalline olivine (Mg$_{(2 −
  2x)}$Fe$_{(2x)}$SiO$_4$). The 69\,$\mu$m band showed that all the
sources produce pure forsterite grains containing no iron in their
lattice structure.  They fit the 69\,$\mu$m band and used its width and
wavelength position to probe the composition and temperature of the
crystalline olivine. The fits showed that on average the temperature of the
crystalline olivine is highest in the group of OH/IR stars and the
post-AGB stars with confirmed Keplerian disks. The temperature is
lower for the other post-AGB stars and lowest for the PNe. A couple of
the detected 69\,$\mu$m bands are broader than those of pure
magnesium-rich crystalline olivine, which can be due to a temperature
gradient in the circumstellar environment of these stars.

\section{HIFI}

\subsection{HIFISTARS}

The Herschel guaranteed time key programme HIFISTARS (\cite[Bujarrabal et
  al. 2012]{Bujarrabal12}) aimed to study the physical conditions,
particularly the excitation state, of the intermediate-temperature gas
in proto-PNe and young PNe.  The information that the observations of
the different components deliver is of particular importance for the
wind-shock interaction and hence the understanding the evolution and
shaping of PNe.  They performed Herschel/HIFI observations of
intermediate-excitation molecular lines in the
far-infrared range of a sample of ten nebulae. The high
spectral resolution provided by HIFI allows the accurate measurement
of the line profiles. The dynamics and evolution of these nebulae are
known to result from the presence of several gas components, notably
fast bipolar outflows and slow shells (that often are the fossil AGB
shells), and the interaction between them. Because of the diverse
kinematic properties of the different components, their emission can
be identified in the line profiles. The observation of these
high-energy transitions allows an accurate study of the excitation
conditions, particularly in the warm gas, which cannot be properly
studied from the low-energy lines.  They detected far infrared lines of
several molecules, in particular of $^{12}$CO, $^{13}$CO, and
H$_2$O. Emission from other species, like NH$_3$, OH, H$_2$$^{18}$O,
HCN, SiO, etc., has been also detected. Wide profiles showing
sometimes spectacular line wings have been found. In the case of
CRL\,618 the $^{12}$CO and $^{13}$CO high excitation line profiles
present a composite structure showing spectacular wings in some cases,
which become dominant as the energy level increases (\cite[Soria-Ruiz
  et al. 2013]{Soria-Ruiz13}). \cite[Bujarrabal et
  al. (2012)]{Bujarrabal12} mainly studied the excitation properties
of the high-velocity emission, which is known to come from fast
bipolar outflows. From comparison with general theoretical
predictions, they find that CRL\,618 showed a particularly warm fast
wind $\sim$300\,K, hotter than previously estimated (\cite[Soria-Ruiz
  et al. 2013]{Soria-Ruiz13}).  In contrast, the fast winds in
OH\,231.8$+$4.2 and NGC\,6302 are cold, T$_k$$\sim$30\,K. Other
nebulae, like CRL\,2688, show intermediate temperatures, with
characteristic values around 100\,K. They argue that the differences
in temperature in the different nebulae can be caused by cooling after
the gas acceleration (that is probably caused by shocks). For
instance, CRL~618 is a case of very recent acceleration of the gas by
shocks, less than $\sim$100\,yr ago, while the fast gas in
OH\,231.8$+$4.2 was accelerated $\sim$1000\,yr ago. They also find
indications that the densest gas tends to be cooler, which may be
explained by the expected increase of the radiative cooling efficiency
with density.  The dense central core of CRL\,618 is characterised
by a very low expansion velocity, $\sim$5\,km\,s$^{-1}$, and a strong
velocity gradient. This component is very likely to be the unaltered
circumstellar layers that are lost in the last AGB phase, where the
ejection velocity is particularly low. The physical properties of the
diffuse halo and the double empty shell, contribute to its line
profiles mainly in the low-J CO transitions (\cite[Soria-Ruiz
  et al. 2013]{Soria-Ruiz13}).

\subsection{Shapemol}

Herschel/HIFI has opened a new window for probing molecular warm gas.
On the other hand, the software {\tt SHAPE} (\cite[Steffen \& Lopez
  2006]{Steffen06}, \cite[Steffen et al. 2011]{Steffen11}) has emerged
in the past few years as a standard tool for determining the
morphology and velocity field of different kinds of gaseous emission
nebulae via spatio-kinematical modelling. {\tt SHAPE} implements
radiative transfer solving, but it is only available for atomic
species and not for molecules.  {\sc Shapemol} (\cite[Santander-García
  et al. 2015)]{Santander-Garcia15} is a complement to {\tt SHAPE}
which enables user-friendly, spatio-kinematic modelling with accurate
non-LTE calculations of excitation and radiative transfer in CO
lines. {\tt Shapemol} is a plug-in completely integrated within {\tt
  SHAPE v5} . It allows radiative transfer solving in the $^{12}$CO
and $^{13}$CO\,J\,$=$1$−$0 to J\,$=$\,17$−$16 lines, but its
implementation permits easily extending the code to different
transitions and other molecular species, either by the code developers
or by the user. Used along {\tt Shape}, {\tt Shapemol} allows easily
generating synthetic maps and synthetic line profiles to match against
observations.

As an example of the power and versatility of {\tt Shapemol}, a model
of the molecular envelope of the planetary nebula NGC\,6302 was made
and compared with $^{12}$CO and $^{13}$CO\,J\,$=$\,2$−$1
interferometric maps from SMA and high-J transitions from HIFI.
\cite[Santander-García et al. (2015)]{Santander-Garcia15} found that
its molecular envelope has a complex, broken ring-like structure with
an inner, hotter region and several “fingers” and high-velocity blobs,
emerging outwards from the plane of the ring. The Herschel spectra are
extremely rich, especially in terms of molecular line transitions.

HIFI data have also allowed a very detailed description of the young
PN NGC~7027. \cite[Santander-Garcia et al. (2012)]{Santander-Garcia12}
also used {\tt Shapemol} doing radiative transfer, spatio-kinematic
modeling of the molecular envelope of the young planetary nebula
NGC~7027 in several high- and low-J $^{12}$CO and $^{13}$CO
transitions observed by HIFI and the IRAM 30\,m radio
telescope, and discussed the structure and dynamics of the molecular
envelope.  They used this code to build a “Russian doll” model to
account for the physical and excitation conditions of the molecular
envelope of NGC~7027.  The model nebula consisted of four nested,
mildly bipolar shells plus a pair of high-velocity blobs. The
innermost shell is the thinnest and showed a significant increase in
physical conditions (temperature, density, abundance, and velocity)
compared to the adjacent shell. This is a clear indication of a shock
front in the system, which may have played a role in shaping the
nebula. Each of the high-velocity blobs is divided into two sections
with considerably different physical conditions. The striking presence
of H$_2$O in NGC~7027, a C-rich nebula, is likely due to photo-induced
chemistry from the hot central star, although formation of water by
shocks cannot be ruled out.

\section{Outlook}

Soon, the {\sc THROES} atlas (Garcia-Lario et al. 2016, this volume)
will be publicly available through the Herschel science archive. {\sc
  THROES} is a catalogue of fully reprocessed, homogenously reduced
PACS spectra of all evolved stars from the AGB to the PN stage,
including some massive red supergiants and LBVs, complemented with
ancillary data taken by other facilities. SPIRE spectra will be added
later. The catalog will contain more than 200 sources, originally part
of more than 40 different research programs.  This will hopefully
trigger additional research by the community.

A lot of Herschel observations of PNe are available to be exploited and
more interesting, scientific results await discovery.

\begin{acknowledgments}

I thank the SOC for inviting me to do this review talk. I am indebted
to my colleagues of the MESS and HerPlanS consortia who have closely
collaborated with her on the Herschel data.  G. Van de Steene wishes
to acknowledge support from FWO through travel grant K1C8716N.  G. Van
de Steene and the MESS consortium wish to acknowledge support from the
Belgian Science Policy office through the ESA PRODEX programme.

\end{acknowledgments}

\end{document}